\documentclass[twocolumn]{aastex61}

\usepackage{amsmath}

\received{1 May 2017}
\accepted{12 June 2017}

\shorttitle{Relativistic Incentive Trap Applied to Starshot}
\shortauthors{R. Heller}

\begin{document}

\title{Relativistic Generalization of the Incentive Trap of Interstellar Travel with Application to Breakthrough Starshot}

\correspondingauthor{Ren\'e Heller}
\email{heller@mps.mpg.de}

\author[0000-0002-9831-0984]{Ren\'e Heller}
\affil{Max Planck Institute for Solar System Research \\
\ Justus-von-Liebig-Weg 3 \\
\ 37077 G\"ottingen, Germany \\
\ \href{mailto:heller@mps.mpg.de}{heller@mps.mpg.de}
}



\begin{abstract}

\noindent
As new concepts of sending interstellar spacecraft to the nearest stars are now being investigated by various research teams, crucial questions about the timing of such a vast financial and labor investment arise. If humanity could build high-speed interstellar lightsails and reach $\alpha$\,Centauri 20\,yr after launch, would it be better to wait a few years, then take advantage of further technology improvements and arrive earlier despite waiting? The risk of being overtaken by a future, faster probe has been described earlier as the incentive trap. Based on 211\,yr of historical data, we find that the speed growth of artificial vehicles, from steam-driven locomotives to Voyager\,1, is much faster than previously believed, about $4.72\,\%$ annually or a doubling every 15\,yr. We derive the mathematical framework to calculate the minimum of the wait time to launch $t$ plus travel time $\tau(t)$ and extend it into the relativistic regime. We show that the $t+\tau(t)$ minimum disappears for nearby targets. There is no use of waiting once we can reach an object within about 20\,yr of travel, irrespective of the actual speed. In terms of speed, the $t+\tau(t)$ minimum for a travel to $\alpha$\,Centauri occurs at $19.6\,\%$ the speed of light ($c$), in agreement with the $20\,\%\,c$ proposed by the Breakthrough Starshot Initiative. If interstellar travel at $20\,\%\,c$ could be achieved within $45$\,yr from today and the kinetic energy be increased at a rate consistent with the historical record, then humans can reach the ten most nearby stars within 100\,yr from today.
\end{abstract}




\section{Introduction}
\label{sec:intro}

The exponential growth of the maximum speed of man-made vehicles, from wind-driven ships, to steam-driven ships and trains, to cars, planes, and space rockets suggests that mankind will reach reasonable speeds for interstellar travel either in this century or the next. The possibility of pushing lightsails to interstellar velocities using Earth-based lasers or the solar photon pressure has in fact been studied as long as half a century ago \citep{1966Natur.211...22M,1967Natur.213..588R,1984JSpRo..21..187F,Macchi2010,macdonald2016advances}. But it has not been until very recently that gram-sized spacecraft have actually been re-considered for interstellar journeys using ultralight photon sails \citep{Lubin2016,2017ApJ...837L..20M,2017ApJ...835L..32H,2017ApJ...834L..20C,2017ApJ...837....5H} with top speeds of up to 20\,\% the speed of light ($c$). Further into the future, fly-bys around nearby stars could use gravity assists \citep{2013JBIS...66..171F} and the stellar photonic pressures \citep{2017arXiv170403871H} to go beyond the solar neighborhood with extremely low demands for on-board propellant.

\subsection{The incentive trap}

The time to reach interstellar targets is potentially larger than a human lifetime, and so the question arises of whether it is currently reasonable to develop the required technology and to launch the probe. Alternatively, one could effectively save time and wait for technological improvements that enable gains in the interstellar travel speed, which could ultimately result in a later launch with an earlier arrival.

Intuitively, one might be inclined to expect that the continuing growth of speed should make it more reasonable for mankind to wait before we set out to the stars, because future spacecraft would be fast enough to overtake any probe that we could send out soon. This conflict has been described as the incentive trap, i.e., the risk of an interstellar space probe to be overtaken by a future probe that has been launched with a velocity high enough to intercept the first probe owing to the ongoing technological progress.\footnote{An early illustration of this scenario was given in the 1944 science fiction short story ``Far Centaurus'' by \citet{Vogt1944}. It pictures a manned mission that takes 500\,yr to reach $\alpha$\,Centauri only to find that the system has already been colonized by humans who actually launched later.} \citet{2006JBIS...59..239K} showed that the total time from now, that is to say, the waiting time to launch $t$ plus the travel time $\tau(t)$, to reach an arbitrary stellar target has a minimum if we assume an exponential growth of the interstellar travel speed $v(t)$. Given the fastest speed of travel at the time (referring to NASA's New Horizon mission), and assuming a $1.4\,\%$ average growth rate in speed, \citet{2006JBIS...59..239K} showed that the minimum of $t+\tau(t)$ to reach Barnard's star, at a distance of about 6\,ly, is 712\,yr from 2006. \citet{2006JBIS...59..239K} also stated that the minimum of the total time will be reached long before relativistic speeds will be achieved.

Here we address the incentive trap under the notion that $v=20\,\%\,c$ can be reached within a few decades from now, as proposed by the Breakthrough Starshot Initiative\footnote{\href{http://breakthroughinitiatives.org/Initiative/3}{http://breakthroughinitiatives.org/Initiative/3}}, or Starshot for short \citep{2017Natur.542...20P}. This would fundamentally change both the assumptions and the implications of the incentive trap because the speed doubling and the compounded annual speed growth laws would collapse as $v$ approaches $c$.

\subsection{Historical speed growth}
\label{sec:historical}

\begin{figure}[t]
\centering
\includegraphics[width=1.0\linewidth]{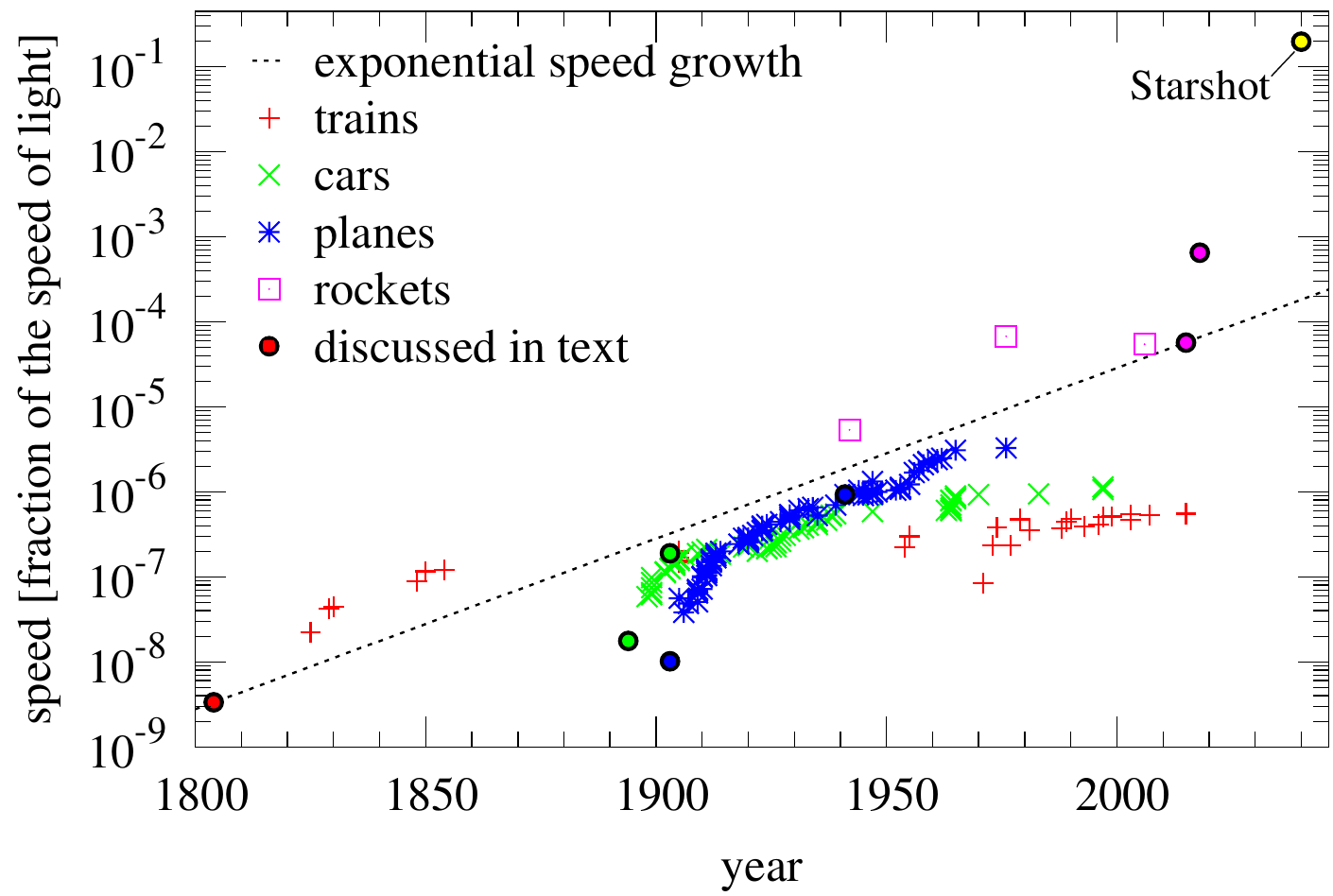}
\caption{Historical speed records of vehicles or probes. Black-rimmed data points are discussed in Section~\ref{sec:historical}. \label{fig:speeds}}
\end{figure}

Let us consider humanity's speed improvements in modern history. The first steam locomotive, Richard Trevithick's ``Penydarren'', made about $4\,{\rm km\,h}^{-1}$ in the year 1804. The world's first production car, the Benz Velocipede of 1894, obtained top speeds of up to $19\,{\rm km\,h}^{-1}$, which was overruled by a $204\,{\rm km\,h}^{-1}$ world record by a Stanley Steamer race car in 1903. At the same time, the top speeds of planes increased from about $11\,{\rm km\,h}^{-1}$ of the Wright Flyer in 1903 to about $1000\,{\rm km\,h}^{-1}$ by the German Messerschmitt rocket-powered planes in the 1940s.

Later on, rockets allowed interplanetary cruises on a timescale of years in the 1960s. In 2015, the Voyager 1 mission has been observed to leave the solar system at a speed of about $17\,{\rm km\,s}^{-1}$ or $5.7\times10^{-5}\,c$ relative to the Sun. And next year, the Solar Probe Plus mission will perform a close solar fly-by with a top heliocentric speed of $195\,{\rm km\,s}^{-1}$ or $6.5\times10^{-4}\,c$  \citep{2016SSRv..204....7F}. The nominal launch of interstellar lightsails with Starshot is in about the year 2040 with a speed of $2\times10^{-1}\,c$.

All these values are symbolized with black-rimmed circles in Figure~\ref{fig:speeds}, with additional top speed measurements of trains, cars, planes, and rockets shown with different symbols (see legend). The dashed black line illustrates an exponential growth law connecting the $1\,{\rm m\,s}^{-1}$ speed of the ``Penydarren'' steam locomotive in 1804 with the $5.7\times10^{-5}\,c$ solar system escape speed of Voyager 1 in 2015. Although this exponential growth captures the development of historic top speeds, we do not claim in this report {\it that} it will continue as such. Instead, we investigate the implications for interstellar travel {\it if} it does continue. Moreover, note the substantial offset of the yellow symbol referring to Starshot. In Section~\ref{sec:discussion} we demonstrate that this jump in velocity in the year 2040 would save about 150\,yr of speed growth according to the historic record.

\section{Analytical models of waiting times and travel times}

\subsection{Doubling laws}

\subsubsection{The speed doubling law}
\label{sec:doubling}

The historical and future speed developments can be addressed with mathematical frameworks, e.g. with an exponential growth of the maximum speed of a vehicle or probe. In its most simple version, the exponential growth law can be written as 

\begin{equation}\label{eq:simple}
v(t) = v_0 2^{t/h} \ ,
\end{equation}

\noindent
where $v_0$ is the current maximum speed, $h$ is the time between doublings, and $t$ is time \citep{2006JBIS...59..239K}. Then $\tau_0=s/v_0$ is the current travel time and $\tau(t)=s/v(t)$ is the future travel time to an object at a distance $s$ from us. With

\begin{align}\label{eq:travel}\nonumber
                                                 & \frac{\tau_0}{\tau(t)} = \frac{s/v_0}{s/v(t)} = 2^{t/h} \\
\Leftrightarrow \hspace{1cm} &\tau(t) \hspace{.1cm} = \frac{\tau_0}{2^{t/h}} \ ,
\end{align}

\noindent
the total time from now to reach the target becomes

\begin{equation}\label{eq:total}
t + \tau(t) = t + \frac{\tau_0}{2^{t/h}} \ .
\end{equation}

\noindent
In this framework, $t + \tau(t)$ depends neither on the distance of the target nor on the current maximum speed. Moreover, relativistic effects are neglected.

The above set of equations has been worked out by \citet{2006JBIS...59..239K} to identify the incentive trap. As an extension to that, we determine the location of the minimum of Equation~\eqref{eq:total}, at a time $t_{\rm min}$ from now, which can be evaluated analytically by means of the derivative. We use the fact that $(u^v)' = u^v (v' \ln(u) + vu'/u)$ for $u>0$, where $u$ and $v$ are functions of $x$ and $u'$ and $v'$ denote their derivatives. Applied to Equation~\eqref{eq:total}, this yields

\begin{align}\nonumber
\frac{d}{dt} {\Big (}t + \tau(t){\Big )} &= 1 + \tau_0 \frac{d}{dt} 2^{-t/h} \\
                                      &= 1 - \frac{\tau_0 \ln(2)}{h} 2^{-t/h} \ ,
\end{align}

\noindent
where $\ln(x)$ symbolizes the natural logarithm of $x$. We then solve

\begin{align}\label{eq:tmin_doub}\nonumber
                                                  \frac{d}{dt} {\Big (}t + \tau(t){\Big )} &= 0 \hspace{2.5cm} , \hspace{1cm} t = t_{\rm min} \\
\Leftrightarrow \hspace{1.5cm} t_{\rm min} &= \frac{\displaystyle - h \ln{\Big (}\frac{h}{\displaystyle \tau_0 \ln(2)}{\Big )}}{\ln(2)} \ .
\end{align}

The historical record (see Section~\ref{sec:historical}) traces a speed growth of about four orders of magnitude over the 211\,yr from the 1804 ``Penydarren'' (about $1\,{\rm m\,s}^{-1}$) to Voyager 1 (about $17,000\,{\rm m\,s}^{-1}$) in 2015 or, alternatively, 14 speed doublings with a speed doubling time of

\begin{align}\label{eq:h}\nonumber
h &= \frac{t}{\log_2(v/v_0)} = \frac{(2015-1804)\,{\rm yr}}{{\Big (} \log_{10}(17,000)/\log_{10}(2) {\Big )}} \\
   &= 15.0\,{\rm yr}
\end{align}

\noindent
(see Section~\ref{sec:discussion} for a discussion of the reference speeds chosen). Plugging this value of $h$ into Equation~\eqref{fig:speeds} yields the black dashed line in Figure~\ref{fig:speeds}. Most important, this value, which is based on the historical record of top speeds is, much smaller than the nominal $100$\,yr assumed by \citet{2006JBIS...59..239K}. As a consequence, we can expect to derive much smaller waiting times to the launch of interstellar probes in our model compared to \citet{2006JBIS...59..239K}.

If this growth can be maintained for another $7\frac{1}{2}$ doublings, i.e. for another about 112\,yr, then humanity would achieve $1\,\%\,c$. That said, the historical record of top speed achievements exhibits jumps whenever new technologies have been introduced. The interstellar speeds aimed at by Starshot could initiate such a speed jump within the next few decades and enable interstellar speeds much earlier.

\begin{figure}[t]
\centering
\includegraphics[width=1.0\linewidth]{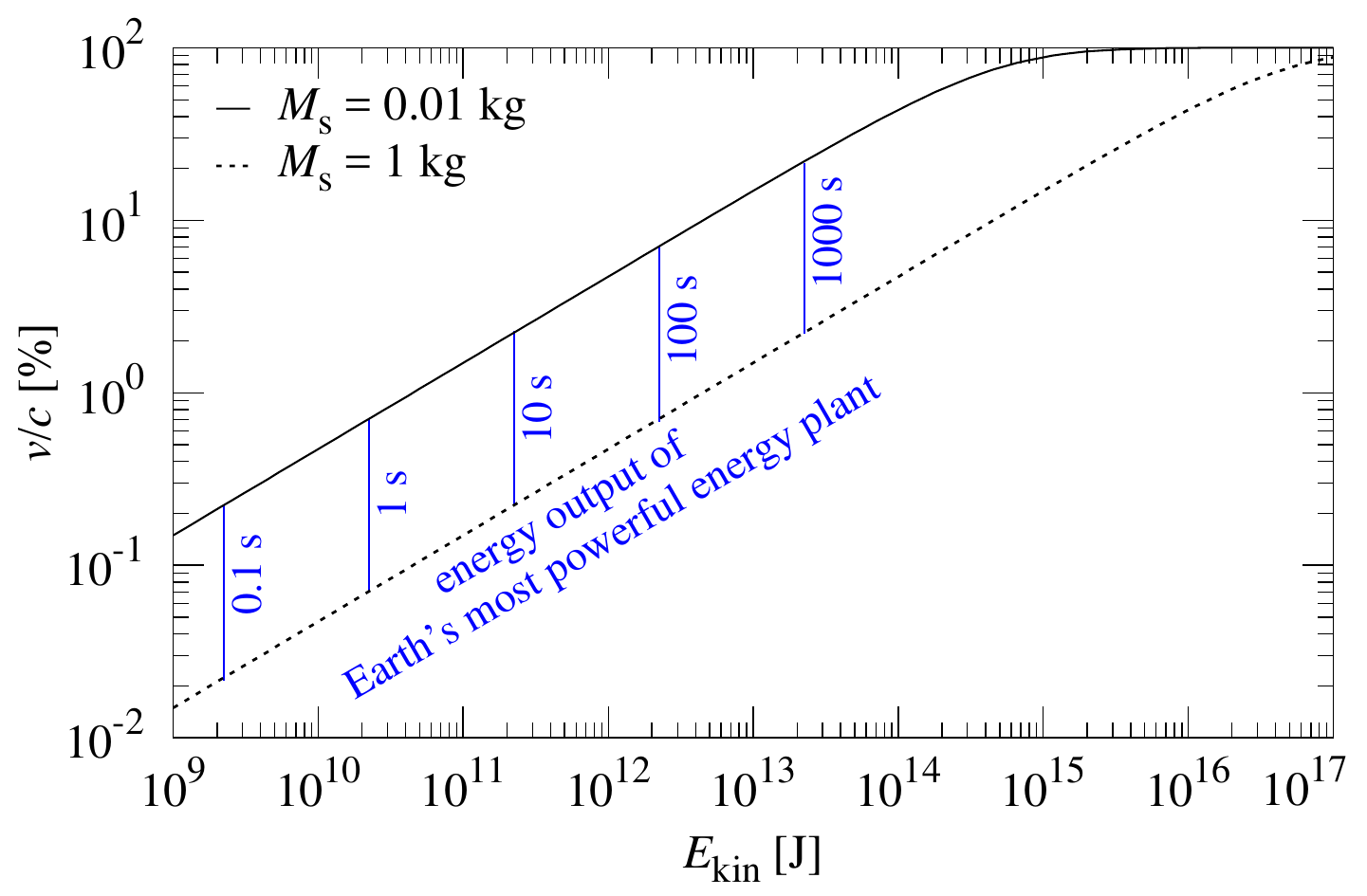}
\caption{Velocity in units of the speed of light as a function of the kinetic energy for two different masses of a hypothetical space probe (Equation~\ref{eq:velocity_rel}). The energy output of the Three Gorges Dam in China, with an installed capacity of 22.5\,GW, is indicated after 0.1\,s to 1000\,s. \label{fig:energy}}
\end{figure}

\subsubsection{The energy growth law}

The speed doubling law is naturally restricted to non-relativistic speeds. Beyond $0.5\,c$, the speed cannot possibly double anymore. We now consider a model, in which the growth of the kinetic energy ($E_{\rm kin}$) is considered instead. If technological progress permits, then $E_{\rm kin}$ could allow relativistic speeds of gram-to-kilogram sized probes within decades from now \citep{Lubin2016}. Figure~\ref{fig:energy} shows the gain in speed for an increase of the kinetic energy pumped into a 1\,kg sail (dashed line) and a 0.01\,kg sail (solid line), the latter of which corresponds roughly to the nominal weight of a Starshot sail. Note that the most powerful power plant on Earth today, the Three Gorges Dam in China, can reach power outputs of up to 22.5\,GW. A 1\,kg (or 0.01\,kg) probe gaining kinetic energy at the same rate for the duration of 100\,s would reach terminal speeds of $0.7\,\%\,c$ (or $7.1\,\%\,c$).

In the non-relativistic regime, Equation~\eqref{eq:simple} is equivalent to

\begin{equation}
v(t) = \sqrt{\frac{2E_{\rm kin,0}}{M_{\rm s}}} 2^{t/h} \ ,
\end{equation}

\noindent
where $M_{\rm s}$ is the rest mass of the space probe and $E_{\rm kin,0}$ is the maximum kinetic energy that can possibly be transferred to the probe today. Hence, the development of the kinetic energy obeys the following law:

\begin{equation}\label{eq:Ekin_doub}
E_{\rm kin}(t) = E_{\rm kin,0} \ 2^{2t/h} \ .
\end{equation}

\noindent
The total energy of the probe then is $E={\gamma}M_{\rm s}c^2$, with $\gamma(v)=(1-(v/c)^2)^{-1/2}$, so that

\begin{equation}
v(t) = c \ \sqrt{1 - {\Big (}\frac{M_{\rm s}c^2}{E(t)}{\Big )}^2} \ .
\end{equation}

\noindent
With $E(t)=M_{\rm s}c^2+E_{\rm kin}(t)$ we have

\begin{equation}\label{eq:velocity_rel}
v(t) = c \ \sqrt{1 - {\Big (}\frac{M_{\rm s}c^2}{M_{\rm s}c^2  + E_{\rm kin,0} \ 2^{2t/h}}{\Big )}^2} \ .
\end{equation}

\noindent
In analogy to Equations~\eqref{eq:travel} and \eqref{eq:total}, we can calculate the future travel time to an object as

\begin{equation}\label{eq:relative_doub}
t + \tau(t) = t + \frac{\tau_0 v_0}{c} \frac{1}{\sqrt{1 - {\Big (}\frac{\displaystyle M_{\rm s}c^2}{\displaystyle M_{\rm s}c^2 + E_{\rm kin,0}2^{2t/h}}{\Big )}^2}} \ ,
\end{equation}

\noindent
which has a derivative

\begin{align}\label{eq:E_rel_der}\nonumber
\frac{d}{dt} {\Big (}t + \tau(t){\Big )} = 1 - \frac{\tau_0 v_0}{hc} \frac{(M_{\rm s}c^2)^2 \, \ln(4) \, 2^{-t/h}}{\sqrt{E_{\rm kin,0}(2M_{\rm s}c^2+E_{\rm kin,0}4^{t/h})^3}} \ . \\
\end{align}

\noindent
The minimum of Equation~\eqref{eq:E_rel_der} will be determined numerically below. Note that the total wait plus travel time in Equation~\eqref{eq:relative_doub} depends on $v_0$, which is different from the speed doubling law described in Equation~\eqref{eq:total}.

\subsection{Compounded growth of speed or energy}

\subsubsection{Compounded growth of speed}

As an alternative to the speed doubling law, \citet{2006JBIS...59..239K} proposed that a compound annual growth law, as it is often used in industry and investment business, could offer a more realistic description of future speed developments:

\begin{equation}\label{eq:compounded}
v(t) = v_0 (1 + r)^{t/{\rm yr}} \ ,
\end{equation}

\begin{figure*}
\centering
\includegraphics[width=1.\linewidth]{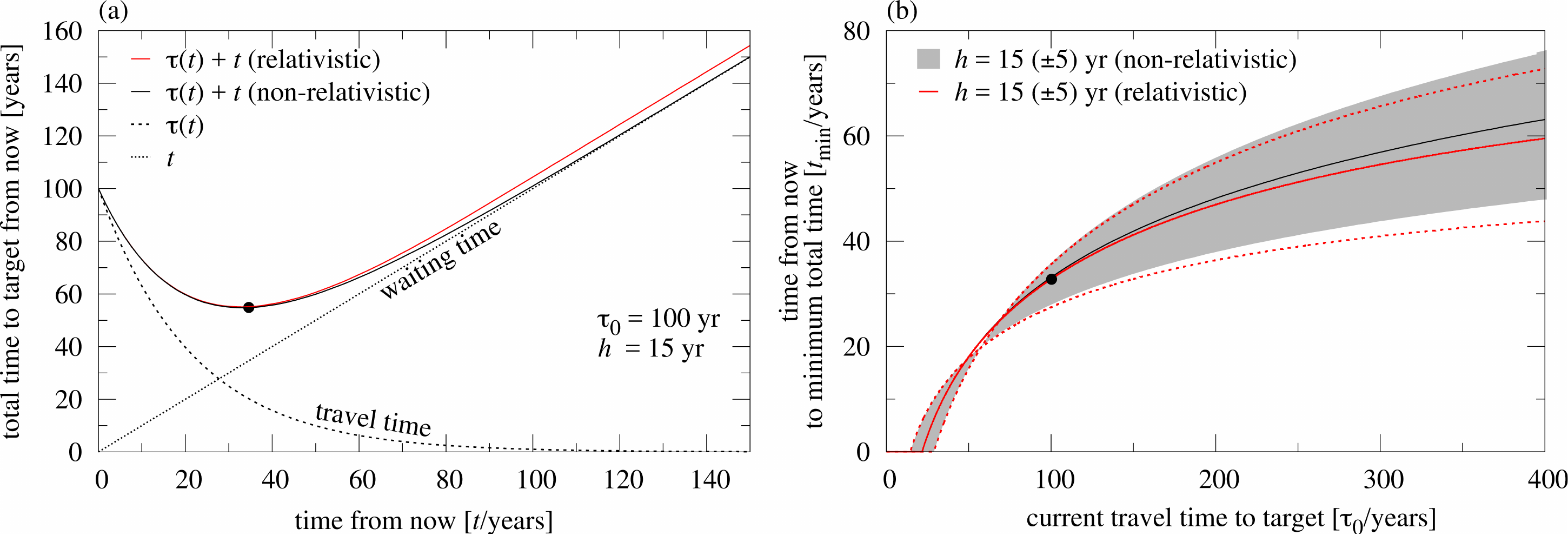}
\caption{Waiting times $t$ and travel times $\tau(t)$ assuming a speed doubling (black lines) or relativistic energy (red lines) growth law. (a) Total time to the target (black solid line) as in the \citet{2006JBIS...59..239K} non-relativistic model, assuming speed doubling after a time $h$ (Equation~\ref{eq:total}). The dashed and dotted lines show the contributions of waiting time and travel time. The red line shows the relativistic correction assuming continuous energy growth (Equation~\ref{eq:relative_doub}). (b) Minimum of the total waiting plus travel time as a function of the current travel time $\tau_0$. The doubling time is taken to be $h=15\,\pm5$\,yr, with the models using the nominal value shown with solid lines. The black line and the grey envelope illustrate the non-relativistic speed doubling (Equation~\ref{eq:tmin_doub}). Red lines visualizes the relativistic energy growth law based on numerical calculations of the minimum of Equation~\eqref{eq:E_rel_der}. \label{fig:totaltime_doubling}}
\end{figure*}

\noindent
where $r$ is the annual percental growth rate.\footnote{Note that the exponent $t$ in Equation~(3) of \citet{2006JBIS...59..239K} must be divided by 1\,yr.} In fact, Equation~\eqref{eq:compounded} is equivalent to Equation~\eqref{eq:simple} for

\begin{equation}\label{eq:t_comp}
r = 2^{{\rm yr}/h} - 1 \ \ \ , \ \ \ h = \frac{\rm yr}{\log_2(1+r)} \ \ \ \ ,
\end{equation}

\noindent
which can be derived by equating Equations~\eqref{eq:simple} and \eqref{eq:compounded} and then solving for either $r$ or $h$.

The total time to arrival is then given as

\begin{equation}\label{eq:t_comp}
t + \tau(t) = \frac{\tau_0}{(1+r)^{t/{\rm yr}}} + t
\end{equation}

\noindent
and we determine the derivative as

\begin{align}\nonumber
\frac{d}{dt} {\Big (}t + \tau(t){\Big )} &= 1 - \frac{\tau_0}{\rm yr} \frac{\ln{(1+r)}}{(1+r)^{t/{\rm yr}}} \ ,
\end{align}

\noindent
the minimum of which is located at

\begin{align}\label{eq:tmin_comp}\nonumber
                                                  \frac{d}{dt} {\Big (}t + \tau(t){\Big )} &= 0 \hspace{2.93cm} , \hspace{1cm} t = t_{\rm min} \\
\Leftrightarrow \hspace{1.5cm} t_{\rm min} &= \frac{ \ln{\Big (}\frac{\displaystyle \tau_0}{\displaystyle \rm yr} \ln(1+r) {\Big )} }{\ln(1+r)} {\rm yr} \ .
\end{align}

The historical speed record outlined in Section~\ref{sec:historical} tracks an annual speed growth of

\begin{equation}\label{eq:r}
r = \left( \frac{v}{v_0} \right)^{{\rm yr}/t} - 1 = \left( 17,000 \right)^{1/211} - 1 = 4.72\,\%
\end{equation}

\noindent
(see black dashed line in Figure~\ref{fig:speeds}), suggesting that humanity will achieve $1\,\%\,c$ within 112\,yr. This value for rhe annual growth rate is substantially larger than the fiducial $1.4\,\%$ assumed by \citet{2006JBIS...59..239K}.

\subsubsection{Compounded growth of energy}

Just like the speed doubling model, the compounded growth of speed becomes impossible near the speed of light, which is why we now address the compounded growth of energy as per
 
 \begin{equation}
 E_{\rm kin}(t) = E_{\rm kin,0}(1+r)^{2t/{\rm yr}} \ ,
 \end{equation}

\noindent
which means that the speed grows as

 \begin{equation}\label{eq:v_comp}
v(t) = c \ \sqrt{ 1 - {\Big (} \frac{\displaystyle M_{\rm s}c^2}{\displaystyle M_{\rm s}c^2 + E_{\rm kin,0}(1+r)^{2t/{\rm yr}}} {\Big )}^2 } \ ,
 \end{equation}

\noindent
so that the total wait plus travel time becomes

\begin{equation}\label{eq:E_comp}
t+ \tau(t) = t + \frac{\tau_0 v_0}{c} \frac{1}{\sqrt{ 1 - {\Big (}  \frac{\displaystyle M_{\rm s}c^2}{\displaystyle M_{\rm s}c^2 + E_{\rm kin,0}(1+r)^{2t/{\rm yr}}} {\Big )}^2}} \ .
\end{equation}

\noindent
The minimum of the derivative

\begin{align}\label{eq:E_comp_der}\nonumber
\frac{d}{dt} {\Big (}t+ \tau(t){\Big )} = \, & 1 - \frac{2 v_0}{c} \frac{\tau_0}{\rm yr} \\ \nonumber
 & \times \frac{(M_{\rm s}c^2)^2 \ln(1+r)(1+r)^{-t/{\rm yr}}}{\sqrt{E_{\rm kin,0}(2M_{\rm s}c^2+E_{\rm kin,0}(1+r)^{2t/{\rm yr}})^3}} \ . \\
\end{align}

\noindent
is calculated numerically in the next section.

\section{Results}
\label{sec:results}

\begin{figure*}[t]
\centering
\includegraphics[width=1.\linewidth]{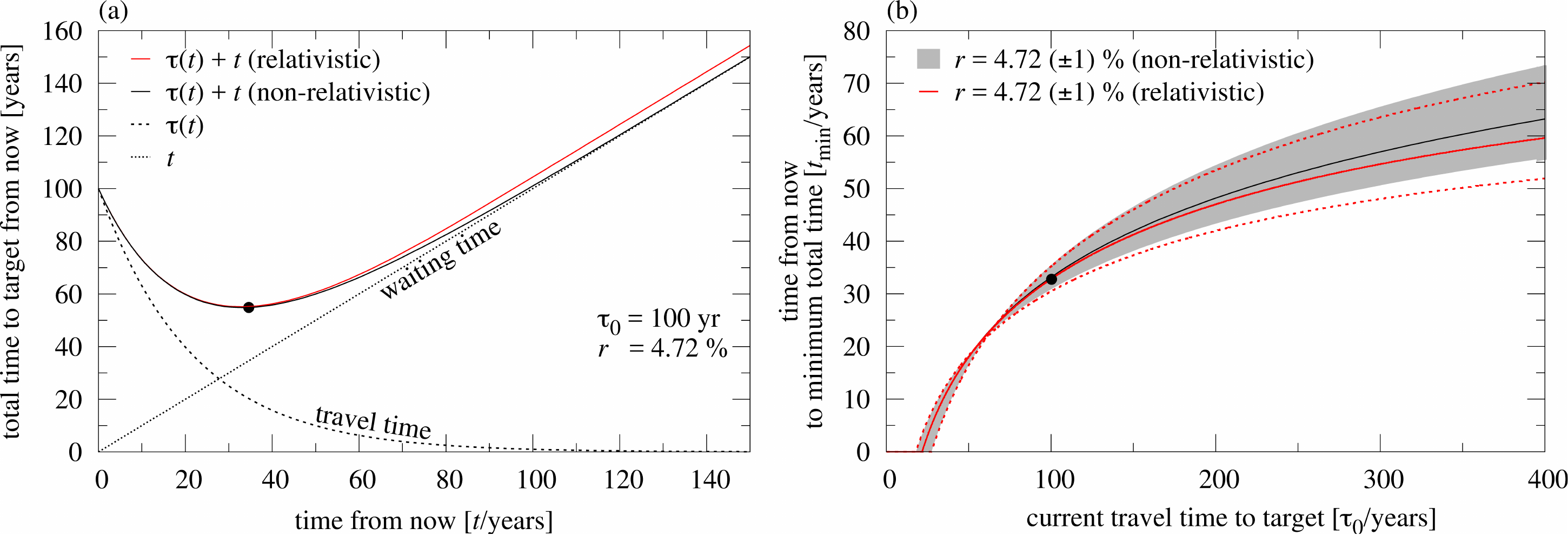}
\caption{Travel times $\tau(t)$ and waiting times $t$ assuming compounded growth of speed (black lines) or energy (red lines). (a) Total time to the target (black solid line) in the non-relativistic model of unlimited compounded speed growth, assuming an annual growth rate of 4.72\,\% (Equation~\ref{eq:t_comp}). The dashed and dotted lines show the contributions of waiting time and travel time. The red line shows the relativistic correction assuming compounded growth of energy (Equation~\ref{eq:E_comp}). (b) Minimum of the total waiting plus travel time as a function of the current travel time $\tau_0$. The black line and the grey strip of 1\,\% tolerance in $r$ illustrate the non-relativistic speed doubling (Equation~\ref{eq:tmin_comp}). Red lines visualize the location of the minima in the relativistic, compounded growth of energy law based on numerical determinations of the minimum of Equation~\eqref{eq:E_comp_der}. \label{fig:totaltime_compound}}
\end{figure*}

\subsection{Speed doublings and relativistic correction}

\subsubsection{The speed doubling law}

The black lines in Figure~\ref{fig:totaltime_doubling}(a) show the different contributions of the travel time (dashed) and the waiting time (dotted) in the non-relativistic approach, assuming a doubling time $h=15$\,yr and $\tau_0=100$\,yr for the travel time, e.g. to a hypothetical nearby star. In this example, the minimum occurs within about $t_{\rm min}=33$\,yr, when the sum of $t$ and $\tau(t_{\rm min})$ is about 55\,yr (see the black circle).

More precisely, Equation~\eqref{eq:tmin_doub} yields $t_{\rm min}=33.1$\,yr. The gain in numerical precision compared to a graphical interpretation in Figure~\ref{fig:totaltime_doubling}(a) is irrelevant given the substantial uncertainties in the growth law, but the formula allows us to conveniently study the expected optimum time for the launch of interstellar probes over a range of possible doubling times and contemporary travel times $\tau_0$.

Figure~\ref{fig:totaltime_doubling}(b) shows the location of the minimum of $t+\tau(t)$ for a doubling time of 15\,yr (black solid line), with a grey strip visualizing a tolerance of $\pm5$\,yr in $h$, as a function of the current travel time $\tau_0$. The black circle on the solid line (for $h=15$\,yr) at $\tau_0=100$\,yr refers to its counterpart in Figure~\ref{fig:totaltime_doubling}(a) at the minimum of $t+\tau(t)$.

An interesting result from this plot is that, taking $h=15$\,yr as a reference, there is no future minimum in $t+\tau(t)$ if the current travel time to the target is smaller than about 22\,yr. For larger doubling times, the minimum disappears at larger values of $\tau_0$ and vice versa. In words, if the travel time to an object is short in the first place, then there is no point in waiting for speed improvements. Any further growth in the maximum possible speed would be negligible due to the short travel time, whereas the waiting time would dominate.

We refer to this minimum value of $\tau_0$ for which the expected speed improvements make it worth waiting as the incentive travel time,

\begin{equation}
\tau_{\rm inc} = \min{\Big (}\tau_0(t_{\rm min}){\Big )} = \tau_0(t_{\rm min}=0) \ .
\end{equation}

\noindent
In the non-relativistic regime of the speed doubling law, we can determine $\tau_{\rm inc}$ by setting $t_{\rm min}=0$ in Equation~\eqref{eq:tmin_doub} and solving for $\tau_0$, thus

\begin{equation}\label{eq:incentive_doub}
\tau_{\rm inc} = \frac{h}{\ln(2)} \ .
\end{equation}

\noindent
For the above-mentioned example (solid line in Figure~\ref{fig:totaltime_doubling}b), we find $\tau_{\rm inc}=21.6$\,yr as the minimum current travel time to the target for which it is worth waiting for speed improvements. In other words, any target that we can reach today within $21.6$\,yr of travel cannot be reached earlier based on a doubling of speed every $15$\,yr.

\subsubsection{Relativistic correction of the speed doubling law}

The red line in Figure~\ref{fig:totaltime_doubling}(a) shows $t+\tau(t)$ in the relativistic regime as per Equation~\eqref{eq:relative_doub}. Different from the non-relativistic model in Equation~\eqref{eq:total}, Equation~\eqref{eq:relative_doub} requires the input of a specific $v_0$ and $M_{\rm s}$ as the doubling of the speed of travel breaks down near the speed of light.\footnote{$E_{\rm kin,0}$ in Equation~\eqref{eq:relative_doub} \, is also determined by $v_0$ and $M_{\rm s}$, that is to say, as per $E_{\rm kin,0}(v_0,M_{\rm s}) = M_{\rm s}c^2 (\gamma(v_0)-1)$.}

For the purpose of illustration, let us consider the travel of a nominal $M_{\rm s}=0.01$\,kg Starshot probe to the $\alpha$\,Centauri ($\alpha$\,Cen) system, which has recently become attractive for interstellar exploration after the detection of an Earth-mass planet in the stellar habitable zone of $\alpha$\,Cen\,C \citep{2016Natur.536..437A}. We adopt a distance of 4.3\,ly to $\alpha$\,Cen and assume $v_0~=~4.3\,\%\,c$ to construct an initial travel time $\tau_0=100$\,yr. The total time to the target agrees in both the non-relativistic model (solid black line) and the relativistic model (solid red line) within 1\,\% up to about 50\,yr from now. Note that ``now'' implies that we could achieve $v_0~=~4.3\,\%\,c$ with the given technology today, which has not actually been demonstrated.

The divergence between the non-relativistic speed doubling and the relativistic model in Figure~\ref{fig:totaltime_doubling}(a) is a manifestation of the fact that the travel time in the relativistic model (not shown) cannot possibly converge to zero but is always restricted to a value $>s/c$, or $>4.3$\,yr for $\alpha$\,Cen. In fact, this value corresponds to the offset between the black solid and the red solid lines.

In particular, while Equation~\eqref{eq:tmin_doub} yields an optimal wait time $t_{\rm min}=33.1$\,yr in the non-relativistic model, our numerical evaluation of Equation~\eqref{eq:E_rel_der} for the derivative of the total wait plus travel time in the relativistic regime yields $t_{\rm min}=32.8$\,yr. The difference in the two models is small since $v_0$ is relatively small compared to $c$ and the distance to the target is also relatively small. Special-relativistic effects kick in for more far away objects (see Section~\ref{sec:nearby}). It is critical to note that the minimum in the relativistic model occurs earlier than in the non-relativistic case. That is because the speed doubling breaks down in the relativistic model.

Knowing $t_{\rm min}$, we can now derive the critical interstellar speed that serves as a benchmark value above which gains in the kinetic energy to be transferred into the probe would not result in smaller wait plus travel times to $\alpha$\,Cen. Assuming $\tau_0=100$\,yr and $v_0~=~4.3\,\%\,c$, a $0.01$ \,kg probe would have a kinetic energy of $E_{\rm kin,0}=M_{\rm s}c^2 (\gamma(v_0)-1)=0.86$\,TJ. We plug $E_{\rm kin,0}~=~0.86$\,TJ and $t_{\rm min}=32.8$\,yr into Equation~\eqref{eq:velocity_rel} and obtain $v(t_{\rm min})~=~19.6\,\%\,c$.

If we consider larger values of $\tau_0$, then $t_{\rm min}$ becomes larger than 32.8\,yr. On the other hand, for a given distance, $v_0=s/\tau_0$ would decrease and so would $E_{\rm kin,0}$. We performed numerical simulations for arbitrary values of $\tau_0$, which show that $v(t_{\rm min})$ is independent of the assumption of $\tau_0$. In other words, whenever humanity will achieve the capability of reaching $v(t_{\rm min})~=~19.6\,\%\,c$, there is no need to wait for speed improvements according to the relativistic correction of the speed doubling law for going to $\alpha$\,Cen. This value is in agreement with the $20\,\%\,c$ proposed by Starshot.

\subsection{Compounded speed growth and relativistic correction}

\subsubsection{Compounded growth of speed}

Figure~\ref{fig:totaltime_compound}(a) shows $t+\tau(t)$ for a compounded speed growth, again assuming $\tau_0=100$\,yr as in Figure~\ref{fig:totaltime_doubling}(a) but now  with an annual speed gain of $r=4.72$\,\%, illustrated by the black line. The grey envelope refers to a variation of $r$ by 1\,\%. Given this parameterization of the model, Equation~\eqref{eq:tmin_comp} yields $t_{\rm min}=33.1$\,yr with a minimum wait plus travel time of 54.8\,yr. 

As in the case of speed doubling, we observe that the minimum of $t+\tau(t)$ disappears for initial travel times $\tau_0 < \tau_{\rm inc}$, the limit of which we determine by setting Equation~\eqref{eq:tmin_comp} equal to zero and by solving for $\tau_0$, thus

\begin{equation}\label{eq:incentive_comp}
\tau_{\rm inc} = \frac{1\,{\rm yr}}{\ln(1+r)} \ .
\end{equation}

\noindent
For our example of $\tau_0=100$\,yr and $r=4.72$\,\% this yields $\tau_{\rm inc} = 21.7$\,yr. Any target that can be reached earlier than that is not worth waiting for further speed improvements in this model.

\subsubsection{Relativistic correction of the law of compounded speed growth}

The red solid line in Figure~\ref{fig:totaltime_compound}(a) denotes the relativistic compounded growth of energy assuming that we could reach $\alpha$\,Cen at $\tau_0=100$\,y. The minimum of the total wait plus travel time occurs at $t_{\rm min}=32.8$\,yr where $t+\tau(t)=55.1$\,yr.

Plugging $E_{\rm kin,0}=0.86$\,TJ and $t_{\rm min}=32.8$\,yr into Equation~\eqref{eq:v_comp} yields $v(t_{\rm min})=19.6\,\%\,c$, in agreement with the $20\,\%\,c$ proposed by Starshot. Again, numerical simulations show that this value is independent of the assumption of $\tau_0$.

The red lines in Figure~\ref{fig:totaltime_compound}(b) refer to the relativistic compounded energy growth law. The red solid line shows $t_{\rm min}$ as a function of $\tau_0$, and the red dotted lines illustrate a variation of $r$ by 1\,\%. Just like in the speed doubling law model or in its relativistic energy growth law, the red lines lie under their non-relativistic counterparts, which is to say that $t_{\rm min}$ occurs earlier if special-relativistic effects are taken into account.

\subsection{Application to nearby stars with Starshot}
\label{sec:nearby}

\begin{figure*}[t]
\centering
\includegraphics[width=1.\linewidth]{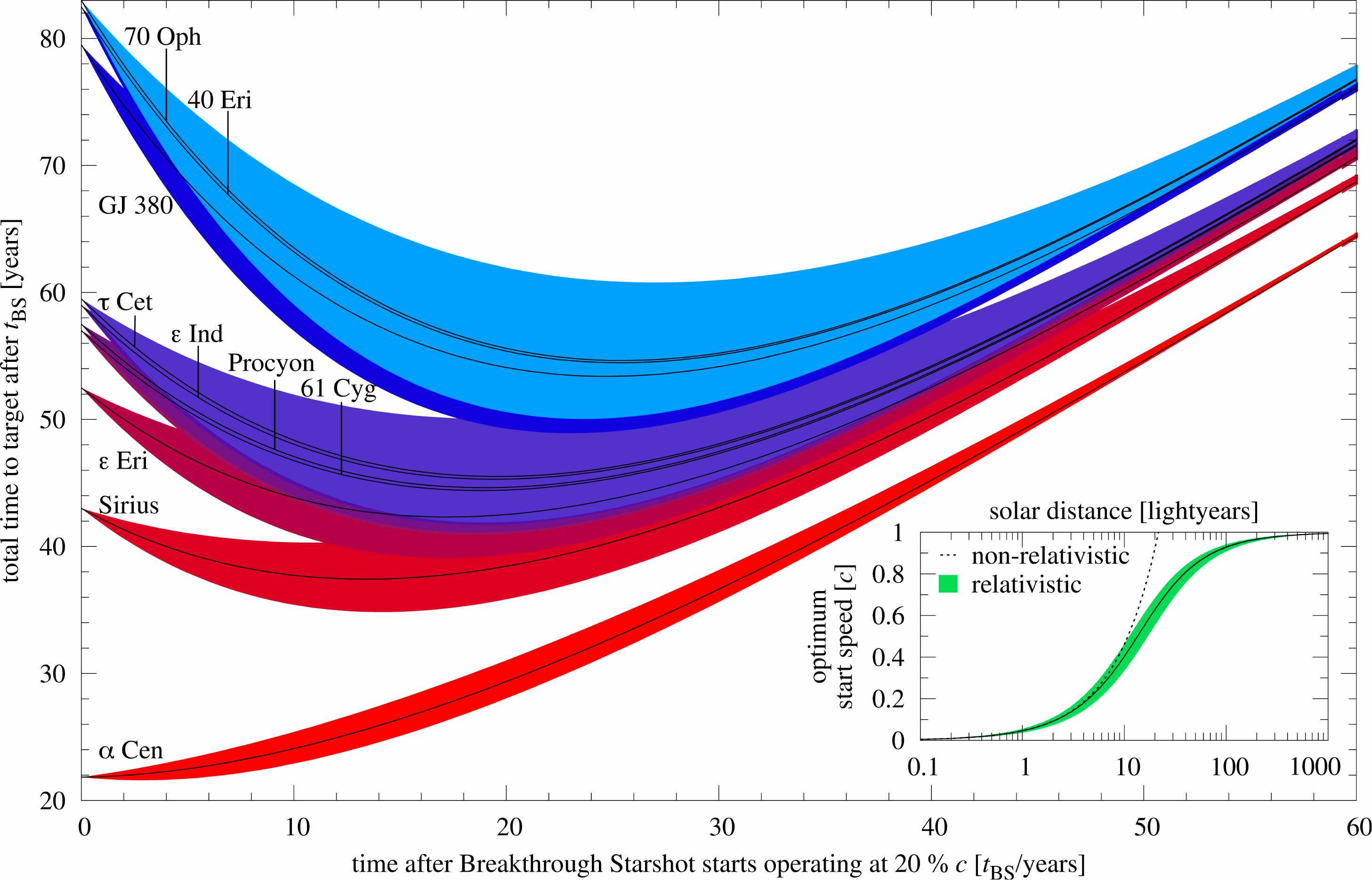}
\caption{Total wait plus travel times to ten of the most nearby stellar systems as a function of time after Starshot starts operating at $20\,\%\,c$. A compounded growth law for the available energy is assumed, with black lines referring to an annual increase in the available energy of $r=4.72\,\%$ based on the last roughly 200\,yr of human speed improvements. The colored strips illustrate $1\,\%$ of tolerance in $r$. The names of the systems are indicated with labels. The inset shows the speed available to humanity at the minimum of the projected wait plus travel time ($v(t_{\rm min})$) as a function of distance to the star. \label{fig:nearby}}
\end{figure*}

\floattable
\begin{deluxetable*}{rcccc}[h!t!]
\tablecaption{Solar distances, optimal launch times after Starshot starts operations, and optimal velocities for minimization of $t+\tau(t)$ to reach ten of the most nearby bright star systems. \label{tab:nearby}}
\tablehead{
\colhead{\#} & \colhead{Name} & \colhead{Distance} & \colhead{$t_{\rm BS}$\tablenotemark{a}}  & \colhead{$v(t_{\rm BS})$\tablenotemark{b}} \\
        &           &     \colhead{(ly)}          &      \colhead{(yr)} & \colhead{\%\,$c$}
}
\startdata
1. & $\alpha$\,Cen\,A/B/C & 4.3 & 0  &  20    \\
2. & Sirius\,A/B                         & 8.6 & 13  &  36    \\
3. & $\epsilon$\,Eri                 & 10.5 & 17  &  42   \\
4. & 61\,Cyg\,A/B                     & 11.4 & 19 &  45   \\
5. & Procyon\,A/B                   &11.5 & 19   &   45   \\
6. & $\epsilon$\,Ind\,A/Ba/Bc & 11.8 & 19  &   46\\
7. & $\tau$\,Cet                      & 11.9 & 20   &   46\\
8. & GJ\,380                            & 15.9 & 25   &   55\\
9. & 40\,Eri\,A/B/C                  & 16.5 & 25   &   57\\
10. & 70\,Oph\,A/B                  & 16.6 & 25  &    57\\
\enddata
\tablenotetext{a}{Minimum of the wait plus travel time after Starshot starts operating at $20\,\%\,c$, see the abscissa in Figure~\ref{fig:nearby}. The values have been determined numerically from the curves in Figure~\ref{fig:nearby}.}
\tablenotetext{b}{Optimal speed for the reduction of $t+\tau(t)$ if improvements of the Starshot technology would permit an annual speed gain by $r=4.72\,\%$.}
\end{deluxetable*}

We now apply the relativistic correction of the compounded growth law to our most nearby stars (see Table~\ref{tab:nearby}), assuming a 0.01\,kg sail and an available travel speed of $v_0~=~20\,\%\,c$ envisioned by Starshot. Although $v_0=20\,\%\,c$ is not accessible for a space probe today, the results of this section inform us about the optimal strategy of the exploration of nearby stars once Starshot goes on line. The relativistic correction of the speed doubling law is not studied in detail, but we confirmed that its predictions do not differ significantly from the ones derived with the relativistic compounded growth law.

In Figure~\ref{fig:nearby} we plot $t+\tau(t)$ for the ten targets as per Equation~\eqref{eq:E_comp}, where we substitute $\tau_0 v_0$ with the stellar distance $s$. Black lines refer to $r~=~4.72\,\%$ and the colored shades illustrate a 1\,\% tolerance. The origin of the coordinate system is chosen at a time $t_{\rm BS}=0$ when Starshot goes into service and allows $v_0=20\,\%\,c$ at the time. Table~\ref{tab:nearby} lists $t_{\rm BS}$ at the minima of $t+\tau(t)$ as well as the corresponding speeds $v(t_{\rm BS})$ for each star. These are the launch speeds beyond which the compounded growth law does not suggest any further reduction in the wait plus travel time.

Figure~\ref{fig:nearby} shows that as soon as interstellar speeds of $20\,\%\,c$ become available, there is no use in waiting for further speed growth to go to $\alpha$\,Cen. Furthermore, the minima of the waiting plus travel times for the next nine systems in our list all occur within 25\,yr or less after the time Starshot will have gone into service, with total times to the target $\lesssim55$\,yr. In other words, if Starshot would become available in as late as 45\,yr from today and if the kinetic energy pumped into $0.01$\,kg probes can be increased at a rate consistent with the historical record of the last two centuries, then humans can reach the ten most nearby bright stars within 100\,yr from today.

The inset in the lower right of Figure~\ref{fig:nearby} shows $v(t_{\rm min})$, the speed (in units of $c$) at the time of the minimum wait plus travel time, as a function of distance out to 1000\,ly. The solid curve assumes an annual speed growth rate of $r=4.72\,\%$. The green strip illustrates a $1\,\%$ tolerance in $r$. For the closest targets, this optimal speed converges to small values, e.g. to $v(t_{\rm min})=0.5\,\%\,c$ for a target at $0.1\,{\rm ly}=6320$\,AU. The function then rises to $v(t_{\rm min})=50\,\%\,c$ at $13.5$\,ly and eventually converges to $c$ for larger distances.

For objects closer than 2\,ly, the non-relativistic model (dashed black line) and the relativist model differ by less than 0.1\,\% in $v(t_{\rm min})$. For targets beyond about 10\,yr, however, the non-relativistic model predicts $v(t_{\rm min})~=~46\,\%\,c$ as an optimal departure speed, whereas the relativistic model suggests $v(t_{\rm min})=40\,\%\,c$, the latter of which takes into account that future gains in the energy transferred to the probe will not result in as much an increase in speed as in the non-relativistic scenario (but rather in increased mass). We performed numerical simulations to validate that all these values of $v(t_{\rm min})$ are independent of the sail's rest mass and only depend on $r$ in the compounded growth law.

\section{Discussion}
\label{sec:discussion}

Free parameters in exponential growth laws are notoriously hard to determine or predict. Small variations in the free parameters, e.g. of $h$ in Equation~\eqref{eq:h} \, or of $r$ in Equation~\eqref{eq:r}, have dramatic effects on the predictions of the respective model. Hence, the numerical values that we derived from historical data shall only be accepted with reservation.

That said, our long data baseline of speed records spans 211\,yr. Variations of the initial $1\,{\rm m\,s}^{-1}$ or the final $17,000\,{\rm m\,s}^{-1}$ by a factor of two over this baseline would result in variations of $h$ by about 1\,yr or of $r$ by about $0.34\,\%$. The colored strips of 5\,yr tolerance in Figure~\ref{fig:totaltime_doubling} and of 1\,\% tolerance in Figure~\ref{fig:totaltime_compound} thus encapsulate the somewhat arbitrary choice of the reference speeds.

Coming back to Figure~\ref{fig:speeds}, we find that the symbol referring to Starshot sits about three orders of magnitude above the exponential speed doubling law in the year 2040. An equivalent transformation of Equation~\eqref{eq:h}, solving it for $t$, suggests that $v(t)=0.2\,c$ would be achieved in the year 2191 if the top speeds would follow the historical records of the past about 200\,yr. This confirms the truly transformative effect that Starshot would have on humanity as a migrating and exploring civilization as a whole.

New velocity regimes impose new physical or  engineering challenges. In our understanding, this is all absorbed in the growth law as illustrated in Figure~\ref{fig:speeds}. As an example, in the early 19th century people were afraid of losing consciousness if travelling faster than running speed ($\approx~20\,{\rm km\,h}^{-1}$). Later on, vibrations that occur when planes approach the sonic barrier were thought to cause fatal damage ($\approx~1200\,{\rm km\,h}^{-1}$) \citep{Portway1940}. And the heat generated during atmospheric re-entry ($\approx~28,000\,{\rm km\,h}^{-1}$) was considered a main challenge for any manned spaceship returning to Earth \citep{heppenheimer2014}. Certainly other challenges will be identified towards relativistic speeds, e.g. the structural integrity and stability during acceleration \citep{2017ApJ...837L..20M}, the interaction of the space probe with the interstellar medium \citep{2017ApJ...837....5H}, or the aiming accuracy towards the target \citep{2017arXiv170403871H}, to name just a few. That said, these might not turn out to be ultimate limits on the maximum possible speed to reach.

\section{Conclusions}

We studied the top speeds obtained by human-made vehicles over the past $211$\,yr, which can well be described by exponential growth laws, and projected the historical data into the future to investigate the possibility of interstellar travel within the next century. According to our estimates, the historical speed growth is much faster than previously believed, about $4.72\,\%$ annually or a doubling every 15 \,yr, from steam-driven locomotives to Voyager\,1.

Surprisingly, we found that the minimum of the wait plus travel time $t+\tau(t)$, previously described by \citet{2006JBIS...59..239K}, disappears for targets that can be reached earlier than a critical travel time, which we refer to as the incentive travel time $\tau_{\rm inc}$. In the non-relativistic domain, $\tau_{\rm inc}$ depends only on the doubling time (see Equation~\ref{eq:incentive_doub}) or, alternatively, on the annual rate of speed growth (see Equation~\ref{eq:incentive_comp}). As an example, for $h=15$\,yr as derived from historical data, $\tau_{\rm inc}$ is about 22\,yr in both the relativistic and the non-relativistic model, i.e., targets that we can reach within about $22$\,yr of travel are not worth waiting for further speed improvements if speed doubles every 15\,yr. The identification of an incentive travel time is irrespective of the parameterization of the underlying law of the speed growth, though its actual value depends, of course.

In terms of the optimal interstellar velocity for launch, the most nearby interstellar target $\alpha$\,Cen will be worthy of sending a space probe as soon as about $20\,\%\,c$ can be achieved because future technological developments will not reduce the travel time by as much as the waiting time increases. This value is in agreement with the $20\,\%\,c$ proposed by Starshot for a journey to $\alpha$\,Cen. We also investigated the speeds beyond which further speed improvements according the historical data would not result in reduced wait plus travel times to ten of the most nearby bright star systems (see Table~\ref{tab:nearby}). It turns out that these speeds, from $20\,\%\,c$ for $\alpha$\,Cen (at 4.3\,ly) to  $57\,\%\,c$ for 70\,Oph (at 16.6\,ly), would all become available within 25\,yr after Starshot will have started operations. These values were derived under the assumption that Starshot or alternative technology would continuously be upgraded according the historical speed record and they are independent of the sail's rest mass.

If Starshot would go on line within the next 45\,yr and if the kinetic energy transferred into the probes can be increased at a rate consistent with the historical speed record of the last 211\,yr, then humans can reach the ten most nearby stars within 100\,yr from today.

\acknowledgments
The author is thankful to Edwin Turner for a challenging discussion that initiated this study, to Michael Hippke for providing helpful feedback and the data used in Figure~\ref{fig:speeds}, and to the Breakthrough Initiatives for an invitation to the Breakthrough Discuss 2017 conference at Stanford University. He also appreciates the comments to a draft version of this manuscript by an anonymous referee. This work was supported in part by the German space agency (Deutsches Zentrum f\"ur Luft- und Raumfahrt) under PLATO grant 50OO1501. This work has made use of NASA's Astrophysics Data System Bibliographic Services.


\bibliographystyle{aasjournal} 
\bibliography{ms}


\end{document}